\documentclass[]{spie}  %>>> use for US letter paper
\usepackage{amsmath,amsfonts,amssymb}
\usepackage{graphicx,aas_macros}
\usepackage[colorlinks=true, allcolors=blue]{hyperref}

\title{Maximising the sensitivity of next generation multi-object spectroscopy: system budget development and design optimizations for the Maunakea Spectroscopic Explorer }

\author[a,b]{Alan W. McConnachie}
\author[a]{Nicolas Flagey}
\author[a]{Kei Szeto}
\author[c]{Shan Mignot}
\author[a]{Alexis Hill}
\author[d]{Pat Hall}
\affil[a]{The Maunakea Spectroscopic Explorer Project Office, 65-1238 Mamalahoa Hwy Kamuela HI 96743 USA}
\affil[b]{NRC Herzberg, Dominion Astrophysical Observatory, 5071 West Saanich Road, Victoria,
British Columbia, Canada}
\affil[c]{GEPI, Observatoire de Paris, PSL Research University, CNRS, Univ. Paris Diderot,
Sorbonne Paris Cit{\'e}, Place Jules Janssen, 92195 Meudon, FRANCE}
\affil[d]{Department of Physics and Astronomy, York University, Toronto, ON M3J 1P3, Canada}

\authorinfo{Further author information: (Send correspondence to A.W.M.)\\A.W.M.: E-mail: mcconnachie@mse.cfht.hawaii.edu, Telephone: 1-808-885-3188}

\begin{document} 
\maketitle

\begin{abstract}
MSE is an 11.25m telescope with a 1.5 sq.deg. field of view. It can simultaneously obtain 3249 spectra at $R=3000$ from $360-1800$nm, and 1083 spectra at $R=40000$ in the optical. Absolutely critical to the scientific success of MSE is to efficiently access the faint Universe. Here, we describe the adopted systems engineering methodology to ensure MSE meets the challenging sensitivity requirements, and how these requirements are partitioned across three budgets, relating to the throughput, noise and fiber injection efficiency. We then describe how the sensitivity of MSE as a system was estimated at the end of Conceptual Design Phase, and how this information was used to revisit the system design in order to meet the sensitivity requirements while maintaining the overall architectural concept of the Observatory. Finally, we present the anticipated sensitivity performance of MSE and describe the key science that these capabilities will enable.
\end{abstract}


\section{INTRODUCTION}
\label{sec:intro}  % \label{} allows reference to this section

The Maunakea Spectroscopic Explorer (MSE) is the only dedicated,
optical and near-infrared, large aperture ($> 10$\,m), multi-object
spectroscopic facility being designed for first light in the
mid-2020s. It is a re-purposing of the Canada-France-Hawaii Telescope,
within an expanded international partnership and upgraded to a larger
aperture.

A key aspect of MSE science is the spectroscopic follow-up of the
plethora of astrophysical objects identified by current and future
photometric and astrometric surveys at a range of wavelengths. This
includes facilities like Gaia, LSST, SKA, Euclid and WFIRST. Here,
there are literally billions of sources, for which exquisite
multiwavelegth data will be available. However, a major missing link
in this network of facilities and data is optical and near-infrared
spectroscopy, and it is this gap that MSE is designed to fill.

The overwhelming majority of all sources identified by these precursor
surveys are faint, and there already exists many MOS spectrographs and
facilities at smaller (4m and less) aperture. MSE must therefore reach
the fainter objects that cannot be accessed with smaller
facilities. Overall system sensitivity is therefore critical. This
paper describes the methodology and design of MSE to meet the
demanding sensitivity requirements, and discusses ways in which the
design is being iterated to ensure the sensitivity of MSE is unrivaled
and enables new science.

At the previous SPIE Astronomical Telescopes and Instrumentation
meeting, the status and progress of the project were detailed in
Ref.~\citenum{murowinski2016} while an overview of the project design
was given in Ref.~\citenum{szeto2016} and the science based
requirements were explained in Ref.~\citenum{mcconnachie2016}. An
update of the project at the end of conceptual design phase is
presented this year in Ref.~\citenum{szeto2018a} with a review of the
instrumentation suite in Ref.~\citenum{szeto2018b}. Other papers
related to MSE are focusing on: the summit facility upgrade
(Ref.~\citenum{bauman2016, bauman2018}), the telescope optical designs
for MSE (Ref.~\citenum{saunders2016}), the telescope structure design
(Ref.~\citenum{murga2018}), the design for the high-resolution
(Ref.~\citenum{zhang2016, zhang2018}) and the low/moderate-resolution
spectrograph (Ref.~\citenum{caillier2018}, the top end assembly
(Ref.~\citenum{mignot2018, hill2018b}), the fiber bundle system
(Ref.~\citenum{venn2018, erickson2018}), the fiber positioners system
(Ref.~\citenum{smedley2018}), the systems budgets architecture and
development (Ref.~\citenum{mignot2016, hill2018}), the observatory
software (Ref.~\citenum{vermeulen2016}), the spectral calibration
(Ref.~\citenum{flagey2016a, mcconnachie2018a}), the throughput
optimization (Ref.~\citenum{flagey2016b, mcconnachie2018b}), the
observing efficiency (Ref.~\citenum{flagey2018b}), and the overall
operations of the facility (Ref.~\citenum{flagey2018a}).

Section~2 describes the science requirements on the system sensitivity
for MSE. Section~3 describes the formalism by which sensitivity
requirements are translated into high level requirements on system
throughput, noise sources, and injection efficiency. Section~4
describes the allocation of values to these three child budgets, and
Section~5 describes the current estimates for these quantities based
on the conceptual design results. Section 6 describes the design
optimisations that are being considered by the Project Office to
reconcile the design with the requirements, and Section 7 discusses
next steps and summarises.

\section{SCIENCE REQUIREMENTS ON SENSITIVITY}

There are three high level science requirements on sensitivity given
in the science requirements document. The sensitivity requirements,
like all science requirements, have been derived from consideration of
the science described in the Detailed Science Case\cite{mcconnachie2016b}, and in particular
the Science Reference Observations. The SROs describe, in considerable
detail, specific and transformational science programs that are
uniquely possible with MSE. Science requirements are defined as the
core set of science capabilities that MSE must have in order to enable
the science described by the SROs.

There are three science requirements relating to sensitivity, one each
for the low, moderate and high resolution mode of the
facility/instrument. They all have the same form and describe the
signal-to-noise ratio that must be obtained for an observation of a
source with a particular intrinsic flux level. The sensitivity science
requirements for the low/moderate/high resolution modes are as
follows:

\begin{itemize}

\item REQ-SRD-034/035/036: Sensitivity at [low/moderate/high] resolution

In the [low/moderate/high] resolution mode, an extracted spectrum from MSE taken in
the observing conditions described below shall have a signal to noise
ratio per resolution element at a given wavelength that is greater
than or equal to [2/2/10] for a 1 hour observation of a point source with a
flux density of $[0.91/1.4/36] \times 10^{-29}$ ergs/sec/cm$^2$/Hz at that
wavelength, for all wavelengths longer than 400nm. Between $370 - 400$\,nm, the SNR shall not be less than [1/1/5] at any wavelength. The
observing conditions in which this requirement shall be met correspond
to a sky brightness of $[20.7/20.7/19.5]$\,mags/sq.arcsec in the V-band at an
airmass of 1.2, and a delivered image quality at that airmass of 0.6
arcseconds full width at half maximum in the r band.
\end{itemize}

MSE is expected to undertake observations of many different sources
with different luminosities, exposure times and observing
conditions. Thus, the characteristics of the observation described in
the science requirement should be considered a ``reference''
observation only, whose sole purpose is to define the sensitivity of
the MSE system.

\section{METHODOLOGY}

\subsection{Overview}

In order to meet the science requirements for sensitivity, it is
necessary to translate the requirement on SNR (sensitivity) as a
function of wavelength for each mode of MSE into requirements
(budgets) on the principal factors that contribute to the SNR
calculation. In Section~\ref{sec:SNR}, we show that these principal
factors are the system throughput, the fiber injection efficiency (IE,
the fraction of light from a point source that is incident on the
focal plane that enters a science fiber) and the noise (specifically,
the random sources of noise that occur during an observation;
systematics errors are not considered part of the noise budget). The
formalism developed in Section~\ref{sec:SNR} also describes the
inter-relation of each of these components with each other (such that
if one changes, the others must change accordingly to ensure the SNR
requirement is still met).

In what follows, we make an intentional distinction between “budget/allocation” and
“estimate”. The former is the top-down requirement which is driven by the science
considerations. The latter is a bottom-up value, which is considered
feasible given our current knowledge of the subsystem conceptual
designs for MSE.  If the latter is very different from the former
(such that the estimate does not meet the allocation), then this
implies a potential risk for the project as it highlights an area
where we might fail to meet the requirements of the Project. Clearly,
it is the responsibility of the Project Office to reconcile any
discrepancies between them in order to achieve the MSE science
goals. If meeting these science goals ultimately prove infeasible,
then a revist of the science requirements, and thus on the scientific
capabilities of MSE, might be required.

Thus, the process the Project Office has followed to maximise the
sensitivity of MSE, and which is described in detail in the proceeding
sections, is as follows:

\begin{enumerate}
\item translate the science requirements on sensitivity into high-level
  budgets
  for injection efficiency, system throughput and noise (following
  Section~\ref{sec:SNR});
\item partition each of these three child budgets into budget allocations for each of the
  relevant subsystems;
\item compare the performance estimates from the conceptual designs of
  all the subsystems to the budget allocations;
\item revisit the budget allocations and/or designs to reconcile the
  allocations and the estimates where necessary;
\item in cases where reconciliation is possible but difficult, a risk
  will be identified by the Project Office that will need to be
  managed;
\item in cases where reconciliation is not possible, the high level
  science requirements may need to be modified. This is a serious step
  to take that requires ultimate approval by the MSE Management Group.
\end{enumerate}

\subsection{SNR Formalism}\label{sec:SNR}
For an astronomical observation, the SNR is defined as

\begin{equation}
SNR=N_{obj}/\sqrt{N_{noise}}
\end{equation}

\noindent where 

\begin{equation}
N_{noise}=N_{obj} + N_{sky} + N_{other}^2
\end{equation}

Here, $N_{obj}$ is the number of counts per resolution element due to the astronomical object being observed, $N_{sky}$ is the number of counts per resolution element due to the sky background, and $N_{other}$ are the counts (or effective counts) due to all other sources of noise (for example, detector read noise and dark current).

For an astronomical source with an intrinsic flux of $F_{obj}$, observed for a time $t$, by a telescope with a collecting aperture of area $S$, operating a fiber spectrograph with a wavelength resolution of $\Delta \lambda$:

\begin{equation}
N_{obj} =  (F_{obj}  \times \Delta \lambda \times t \times S \times E) / P
\end{equation}

\noindent $P=hc/\lambda$ is the energy per photon, where $h$ is Plank's constant and $c$ is the speed of light. $E$ is the overall efficiency (throughput) of the entire astronomical system including atmosphere. We choose to deconstruct $E$ as follows: 

\begin{equation}
E = E_{atm} \times E_{inj} \times E_s
\end{equation}

$E_{atm}$ is the throughput of the atmosphere. The product of $(E_{inj} \times E_s)$ is the overall throughput of the telescope and instrument (from primary mirror to detectors). That is, it describes how much of the flux from the astronomical target that enters the telescope (is incident on M1) is measured by the instrument detectors. 

Excluding the atmosphere, we have chosen to break the throughput into
two components. The first component is the injection efficiency,
$E_{inj}$, which is the fraction of light from the astronomical source
incident on the focal plane of the telescope that makes it into the
fiber. The second term, $E_s$, refers to the combined throughput of
all other components of the telescope and instrument.

The reason that we have chosen to define the throughput in this way is clear when we consider the number of sky photons detected: 

\begin{equation}
N_{sky} =  (F_{sky}  \times \frac{\pi d^2}{4}  \times \Delta \lambda \times t \times S \times E_s) / P
\end{equation}

Here, $d$ is the diameter of the fiber. Adopting these definitions, $E_s$ can be used in the definition of both the object counts and the sky counts. 

Substituting everything into Equation (2), and making explicit the dependency on wavelength, gives the following: 

{\scriptsize
\begin{eqnarray}
SNR(\lambda)= \left( \left(F_{obj}(\lambda)  \times \Delta \lambda \times t \times S \times E_{atm}(\lambda) \times E_{inj}(\lambda) \times E_s(\lambda)\right) / P(\lambda) \right)/ \nonumber \\ 
\sqrt{ (F_{obj}(\lambda)  \times \Delta \lambda \times t \times S \times E_{atm}(\lambda) \times E_{inj}(\lambda) \times E_s(\lambda)) / P(\lambda)  + (F_{sky}(\lambda)  \times \frac{\pi d^2}{4}  \times \Delta \lambda \times t \times S \times E_s(\lambda)) / P  + N_{other}(\lambda)^2}
\end{eqnarray}
}

\begin{figure} 
   \begin{center}
   \includegraphics[angle=90, width=15cm]{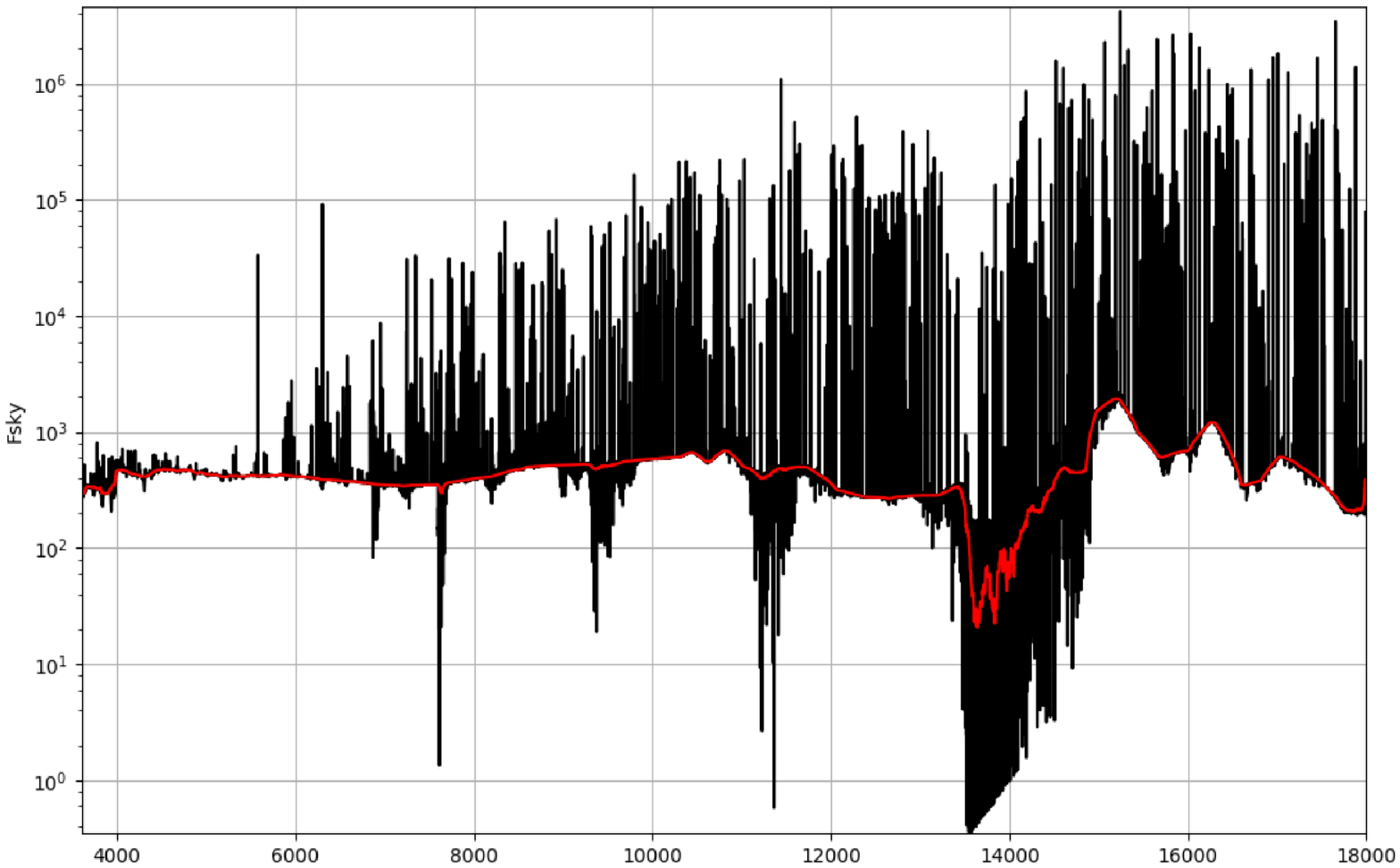}
   \end{center}
   \caption[example] 
%>>>> use \label inside caption to get Fig. number with \ref{}
   { \label{fig:sky} 
Example of the sky spectrum used in sensitivity (SNR) related calculations. Specifically, we use the median-filtered version of the spectrum shown as the red line that traces the continuum level, and which does not include the many strong emission lines.}
   \end{figure}

All relevant terms in Equation (6), except $E_s$, $E_{inj}$ and
$N_{other}$ are defined in the science requirements of MSE or through
the design choices of the Observatory described in the MSE Observatory
Architecture Document.  The values of all known parameters in Equation 6 are summarised in
Table~\ref{params}. We note the following:

\begin{itemize}
\item The object flux levels correspond to sources with monochromatic
  magnitudes of $m= 24.0 / 23.5 / 20$ for the low resolution, moderate
  resolution and high resolution modes, respectively;
\item We use the ESO SkyCalc\cite{noll2012,jones2013} model values,
  modified for Maunakea, for the sky emission spectrum and the
  atmospheric absorption. We use a highly smoothed version of the
  simulated sky spectrum to avoid the influence of the many strong
  emission lines, as shown by the red line in Figure~\ref{fig:sky}.
\end{itemize}

Thus, to meet the overall sensitivity requirements of MSE requires
partitioning appropriate values to the system throughput, injection
efficiency and noise terms of Equation (6) given the adopted values of
all the other parameters described in Table~1.

\begin{table}[ht]
\caption{Summary of all relevant terms for necessary for the calculation of the SNR, as described by Equation (6)} 
\label{params}
\begin{center}   

{\scriptsize      
\begin{tabular}{|p{0.25\textwidth}|l|p{0.3\textwidth}|}
\hline
\rule[-1ex]{0pt}{3.5ex}  Parameter & Value &Notes  \\
\hline
  \rule[-1ex]{0pt}{3.5ex}  $SNR(\lambda)$, Signal-to-noise ratio
                                   & \begin{tabular}{l}LR: $\le
                                       400$nm -- 1 / $>
                                       400$nm -- 2  \\MR: $\le
                                       400$nm -- 1 / $>
                                       400$nm -- 2 \\ HR: $\le
                                       400$nm -- 5 / $>
                                       400$nm -- 10 \end{tabular} & Reference values only \\
\hline
\rule[-1ex]{0pt}{3.5ex}  $F_{obj}(\lambda)$, Flux from object & \begin{tabular}{l}LR: $0.91 \times 10^{-29}$ ergs/sec/cm$^2$/Hz\\MR: $1.4 \times 10^{-29}$ ergs/sec/cm$^2$/Hz\\HR: $36 \times 10^{-29}$ ergs/sec/cm$^2$/Hz\end{tabular} & Reference values only   \\
\hline
\rule[-1ex]{0pt}{3.5ex}  $F_{sky}(\lambda)$, Flux from sky & \begin{tabular}{l}ESO
                                                             SkyCalc
                                                             model
                                                             normalised
                                           to:\\LR: V = 20.7
                                                               mags/sq.arcsec\\MR:
                                                               V =
                                                               20.7
                                                               mags/sq.arcsec\\HR:
                                                               V =
                                                               19.5
                                                               mags/sq.arcsec\end{tabular}&
                                                                                            Modified
                                                                                            for
                                                                                            Maunakea;
                                                                                            Reference
                                                                                            value only \\
\hline
\rule[-1ex]{0pt}{3.5ex}  $E_{atm}(\lambda)$, atmospheric extinction & ESO
                                                             SkyCalc
                                                             model,
                                                                      airmass
                                                                      1.2&
                                                                           Modified
                                                                           for
  Maunakea\\
\hline
\rule[-1ex]{0pt}{3.5ex}  $t$, exposure time (reference) & 3600s & Reference value only \\
\hline
\rule[-1ex]{0pt}{3.5ex}  $S$, collecting area of M1 & 80.8m$^2$& Design choice\\
\hline
\rule[-1ex]{0pt}{3.5ex}  $\Delta\lambda = \frac{\lambda}{R}$, wavelength resolution of
  spectrograph & \begin{tabular}{l}LR: $R=3000$\\MR: $R=6000$ \\HR:
                   $\le 500$nm -- R=40000 /\\HR: $> 500$nm -- R=20000  \end{tabular}& Set by science requirements on resolution \\
\hline
\rule[-1ex]{0pt}{3.5ex}  $d$, diameter of fiber
                                   & \begin{tabular}{l}LR: 1.0
                                       arcsecs\\ MR: 1.0 arcsecs \\
                                                  HR: 0.8
                                       arcsec\end{tabular} &
                                                                 Design
                                                             choice
                                                             based on optimal
                                                                 values
  given expected image quality\\
\hline
\rule[-1ex]{0pt}{3.5ex}  $P(\lambda)$, energy of photon &
                                                          $\frac{hc}{\lambda}$
                                           & Fundamental physics \\
\hline 
\end{tabular}}
\end{center}
\end{table}

\section{DEVELOPMENT OF SENSITIVITY BUDGETS}

The formalism developed in the previous section allows us trade
performance in the areas of system throughput, noise and IE against
each other to meet the overall sensitivity requirements. As the design
matures and our knowledge of the system performance improves, this
three-way trade will be a critical tool at the disposal of the Project
Office to ensure MSE meets its demanding sensitivity requirements.

Prior to having all information available on all relevant aspects of
the MSE performance, we choose to construct the first versions of the
three child budgets (IE, throughput and noise) in the following way:

First, we estimate the expected IE by conducting detailed modeling
based on existing design information as well as engineering
experience. We directly adopt these estimates as the allocation to the
IE budget. Our rationale for doing this is that we expect that there
is limited design alternatives available that can change the IE
significantly, given that we are already pushing extremely hard on the
IE to make it a maximum.

We then estimate all the known sources of subsystem noise, based on
demonstrated performance of components, existing design information,
and engineering experience. Some of these sources of noise depend upon
the final signal (e.g., the expected diffuse light level is a fraction
of the average signal), and so we note that their absolute values
depend upon the overall throughput (including IE), which is initially
unknown. These values (either absolute noise levels or noise levels
relative to the signal) are then allocated to the noise budget.

Finally, we use a custom code to solve the set of equations that we
describe in Section 3 to determine what the throughput budget
allocation must be in order to meet the SNR requirements described in
the SRD. This involves an iterative process whereby the absolute
values of the noise budgets are revised each time the throughput is
calculated, and this process is repeated until convergence.

As the previous section makes clear, the three child budgets of
sensitivity are not independent, and so at the very least some
elements of each of the budgets would have to be calculated - in
contrast to being directly allocated - in order to ensure consistency
with the driving science requirements. In this first version of these
budgets, we have chosen to calculate, rather than independently
allocate, the system throughput. Our rationale for this is that we
expect there to be a bigger design-space to explore to find suitable
solutions that can increase system throughput than there is to
increase injection efficiency and to minimise most sources of noise
compared to our baseline assumptions on these quantities (see next
section). Even if this rationale is wrong, the formalism outlined in
the previous section provides a clear way for us to rebalance the
budgets once we have new information on the expected performance of
the various subsystems of MSE.

Table~\ref{budgets} presents the initial allocations to the injection efficiency,
noise and system throughput budgets calculated in the way described
above.
\begin{table}
\caption{Initial budget allocations to Injection Efficiency (\%
  throughput), Noise (electrons per resolution element) and System
  Throughput (\%).} 
\label{budgets}
\begin{center}     

{\scriptsize  
\begin{tabular}{|l|c|lllllllllllllll|}
\hline
\rule[-1ex]{0pt}{3.5ex}  & $\lambda$, nm &360 & 370 & 400 & 482 &626 &
  767 & 900 & 910 & 950 & 962 & 1235 & 1300 & 1500 & 1662 & 1800\\
\hline
  \rule[-1ex]{0pt}{3.5ex}Injection & LR & 66.4 &67.8&70.4&73.0&74.9&75.6&76.0&76.0&76.1&76.1&76.1&75.8&74.8&73.2&71.9\\
efficiency&MR &66.4&67.8&70.4&73.0&74.9&75.6&76.0&76.0&76.1&&&&&&\\
(\%)&HR & 51.8&53.3&56.0&58.7&61.1&62.1&62.6&&&&&&&&\\
\hline
Noise & LR & 15&15&15&17&17&17&20&20&20&88&88&88&105&89&93\\
($e^-$ / res.& MR& 14&14&15&15&15&15&17&17&17&&&&&&\\
element) & HR &17&17&17&18&18&18&18&&&&&&&&\\
\hline
System&LR &5&4&4&12&10&10&15&15&17&29&24&24&81&30&82\\
Thruput&MR&5&5&5&11&10&10&14&14&16&&&&&&\\
(\%)&HR&10&10&9&21&8&7&7&&&&&&&&\\
\hline
\end{tabular}}
\end{center}
\end{table}
\section{ESTIMATION OF REALIZED PERFORMANCE}

The major system elements that are identified of importance for the
sensitivity of MSE are the following  (considering successive
``Product Breakdown Structure'' (PBS) elements along the optical path):

\begin{enumerate}
\item MSE.ENCL: the enclosure
\item MSE.TEL.STR: the telescope structure
\item MSE.TEL.M1: the primary mirror
\item MSE.TEL.PFHS: the prime focus hexapod system
\item MSE.TEL.WFC/ADC: the wide field corrector and atmospheric dispersion corrector
\item MSE.TEL.InRo: the instrument rotator
\item MSE.SIP.PosS: the fibre positioner system
\item MSE.SIP.FiTS: the fibre transmission system 
\item MSE.SIP.LMR or MSE.SIP.HR: the spectrographs, either
  low/moderate (these two modes are realised in a single spectrograph system) or high resolution
\end{enumerate}

All PBS elements up to and including the instrument rotator are in
common for all three resolution modes.  For MSE.SIP.PosS, the low and
moderate resolution modes share the same fibers/positioners, different
to those of the high resolution mode. For the remaining elements of
the MSE.SIP branch (the ``Science Instrument Package''), the
subsystems are different for each resolution mode of
MSE. Specifically, two different fiber systems lead to two different
spectrograph locations. A $\sim30$\,m fiber cable leads to the
telescope platforms, and feeds into the low/moderate spectrographs. A
seperate $\sim50$\,m fiber cable leads to the coude room in the
telescope pier. This stable environment houses the high resolution
spectrograph suite.

Each of these PBS elements have associated throughputs (e.g.,
vignetting effects, reflectivities, transmissions, etc), and contribute
to the overall system throughput. 

Contributions to the noise are made in the spectrographs, especially
the detector system but elsewhere too (e.g., read noise, dark current,
thermal background, cross-talk, ghosting, instrumental stray
light). Telescope stray light is also a consideration, and directly
affects TEL.STR. Point spread function
considerations also impact the noise estimates (especially via
cross-talk).

The IE budget is discussed in detail in Ref.~\citenum{flagey2018c}. A
critical budget that is related to the IE budget is the Image Quality
budget, that describes the expected contributions to the IQ as
measured at the focal plane of MSE: bigger images due to poorer IQ
mean that less light is able to enter the fiber. In the first version
of these budgets, we assumed that the delivered IQ was 0.6 arcsecs in
the r-band, as per the sensitivity science requirement. However,
subsequent work has shown that the median IQ we can expect is
significantly better than this, and we discuss the impact of this
improvement in the next section.

In 2017, conceptual design studies were completed for all the major
subsystems (10 different reviews for 8 different subsystems). As part
of these studies, throughputs for all major subsystems were estimated
based on actual design information, and a much more detailed
understanding of the system - include noise effects - was able to be
provided than was available in earlier phases. These estimates were
combined via Equation 6 to determine an estimate for the sensitivity
of MSE, to compare to the science requirement described in Section 2.

   \begin{figure} [ht]
   \begin{center}
   \includegraphics[width=10cm]{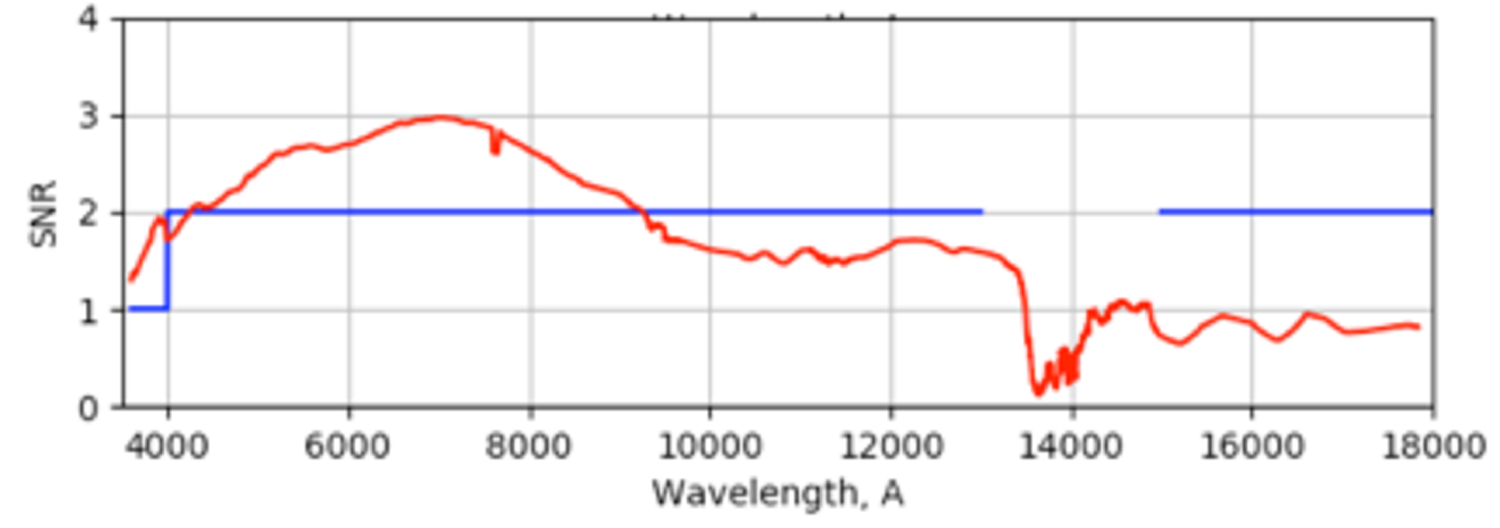}
   \includegraphics[width=10cm]{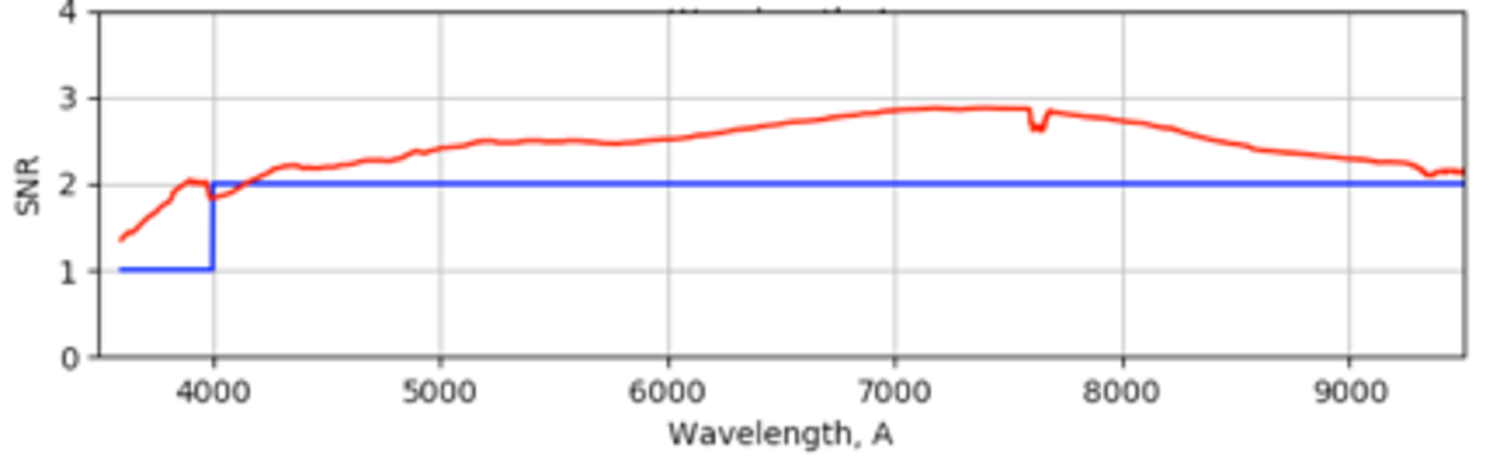}
   \includegraphics[width=10cm]{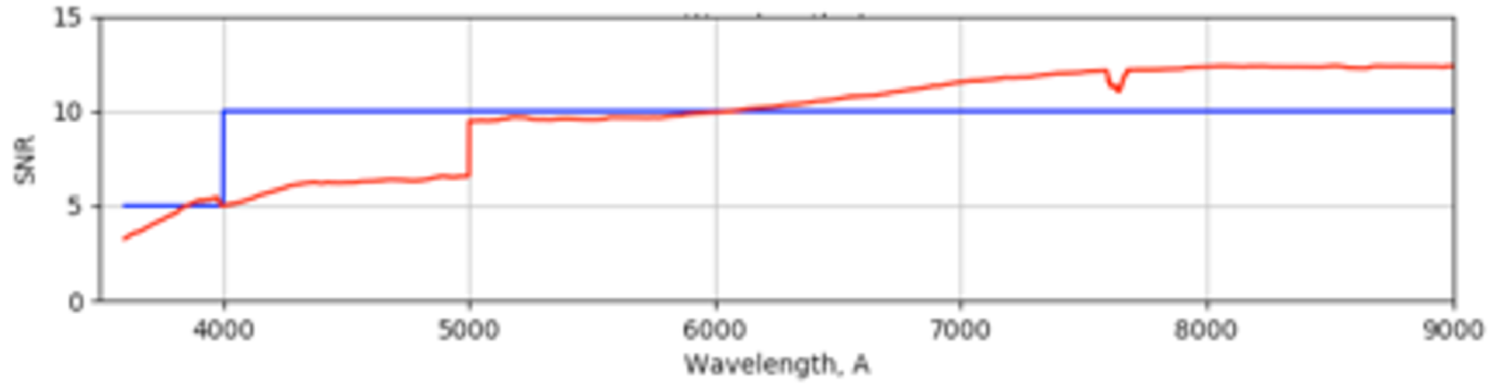}
   \end{center}
   \caption[example] 
%>>>> use \label inside caption to get Fig. number with \ref{}
   { \label{fig:example} 
Initial estimates at the end of Conceptual Design for the sensitivity
of the low (top panel), moderate (middle panel) and high (bottom panel) resolution modes of MSE. Blue is the requirement, red is the estimated performance.}
   \end{figure} 

Figure~2 shows the estimates (red lines) of the sensitivity of MSE as
a function of wavelength compared to the requirements (blue lines) for
the low, moderate and high resolution modes of MSE (top, middle and
bottom panels, respectively). The key points to note here are that:

\begin{itemize} 
\item Low and moderate resolution meet science requirements in the optical except in the $4000 - 4500$A region
\item NIR performance in the low resolution mode is below
  requirements, significantly so for the H-band;
\item The high resolution performance in the blue ($<5000$A)
  does not meet requirements. 
\end{itemize}

\section{DESIGN OPTIMIZATION}

We now discuss the effect of changes made in either the analysis
procedure, or proposed changes in the engineering design of MSE, that can
improve the sensitivity performance of MSE. We
note that in each following subsection, the cumulative effect of all
the recommendations in the preceding subsections are considered.

\subsection{M1: Zecoat}
 
The relatively poor performance in the blue, especially for the HR
mode, motivates examination of adopting a blue-enhanced coating for
M1. Here, we consider the Zecoat coating as a potential candidate. The
Zecoat coating improves performance significantly in the blue. It
ensures we will safely meet science requirements in the low and
moderate resolution modes around 4000A, and has essentially no affect
at longer ($>5000$A) wavelengths. Scientifically, the use of a
Zecoat-like coating for M1 seems to be highly desireable.
 
\subsection{LMR detector operating temperature}

For the low resolution mode at NIR wavelengths, the major source of
subsystem noise (as opposed to sky or object noise) is dark
current. Significantly reducing the dark current will increase the
sensitivity of MSE in the NIR. Our initial estimate of the sensitivity
assumed a dark current of $0.05$e/s/pix (H2RG@153K). However, the dark
current can be significantly lower if the detector operates at a
cooler temperature. Indeed, we estimate that an operating temperature of 95K will give a dark
current of $0.02$e/s/pix, more than half of what we originally
adopted. An operating temperature of 75K is even more desireable,
since here we can anticipate a dark current of only $0.005$e/s/pix or
lower. By adopting a dark current of 0.02e/s/pix, we now expect to
meet the low resolution science requirement in the J band.

\subsection{FiTS/Spectrograph locations}

Given the preciousness of blue photons in the HR mode, we are
considering changing the locations of the LMR and HR, so that the HR
mode can use shorter fibers (which are a major light loss especially
at $< 5000$A). This comes at the expense of blue photons in the low
and moderate resolution modes. However, it is a science requirement
for the HR mode that two out of the three spectral windows that it
will access are at $\lambda < 5000$A. This is due to the rich number
of spectral diagnostics and chemical species that have features at
blue wavelengths. In this respect, a majority of the ``science
content'' of the HR mode is at short wavelengths, more so than in the
low resolution mode.

Our estimate on the throughput changes resulting from this switch in
spectrograph locations is such that we still expect to meet the
sensitivity requirement in the optical for low and moderate resolution
at the blue end because of the previous changes we have incorporated
(especially, the M1 ZeCoat). However, there is a notable increase in
the throughput of HR at the bluest wavelengths (at 3700A, there is a
27\% gain in throughput; at 4000A, there is a 13\% gain). There is
essentially no change at red wavelengths.

\subsection{Fiber injection}
 
For a given IQ, there is a theoretical best IE that can be
achieved. The delivered IE is estimated from the IE budget, and we
would like it to be as close to the theoretical best as
possible. Generally, our modeling of the IE suggests we are around
90\% of the theoretical best.

Our inital examination of this issue assumed a delivered IQ of 0.6
arcsecs in the r-band, as per the science requirements in Section
2. However, results at the end of conceptual design demonstrate that we
can expect a median IQ closer to 0.5 arcsecs, not 0.6 arcsecs. With
this dramatically smaller IQ, the IE for a given fiber size increases
significantly. Specifically, requiring the $IQ=0.5$ instead of 0.6
essentially increases the injection efficiency by $20 - 25$\% at all
wavelengths.

Clearly, pushing further on the IQ is hard. The items that contribute
to it are less well understood and the uncertainty at the end of
conceptual design is larger than for other budgets. However, the gain
in improving the IQ by 0.1 arcseconds is immense.

\subsection{Updated sensitivity estimates}

   \begin{figure} [ht]
   \begin{center}
   \includegraphics[width=10cm]{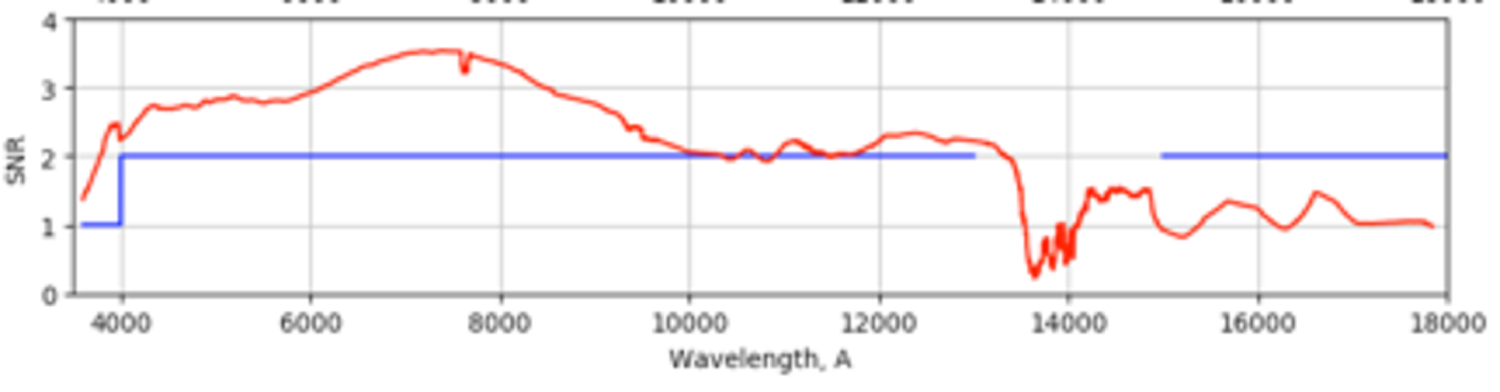}
   \includegraphics[width=10cm]{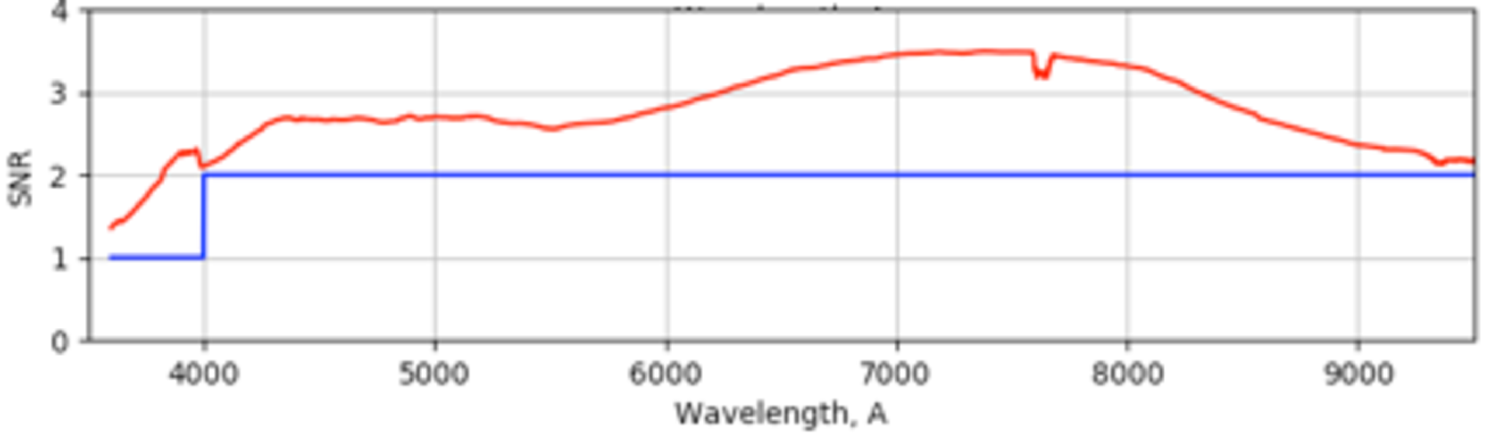}
   \includegraphics[width=10cm]{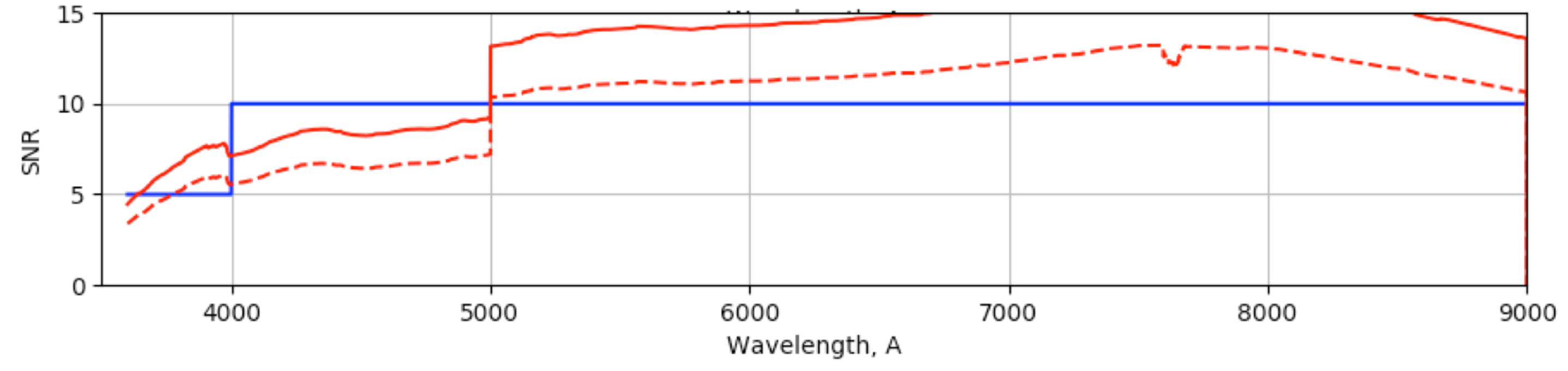}
   \end{center}
   \caption[example] 
%>>>> use \label inside caption to get Fig. number with \ref{}
   { \label{new_est} 
Updated estimates for the sensitivity
of the low (top panel), moderate (middle panel) and high (bottom
panel) resolution modes of MSE, after adopting the optimisations
discussed in the text. Blue is the requirement, red is the estimated
performance. For the high resolution mode, the solid line corresponds
to the SNR level measured at the peak of the grating efficiency ($\sim
80\%$), and
the dashed line corresponds to the SNR level measured at the minimum
grating efficiency ($\sim
50\%$).}
   \end{figure} 

   Figure~\ref{new_est} shows the updated estimates of the sensitivity
   performance of MSE after adopting the optimisations discussed
   previously (M1 coating, LMR operating temperature, spectrograph
   locations and fiber injections. We now meet the science
   requirements everywhere in the low and moderate resolutions except
   for the H-band. For the high resolution mode, two estimates are
   presented, corresponding to the peak throughput of the grating
   (solid line) and the minimum throughput (dashed line). Again, we
   meet science requirements everywhere except in the critical
   $4000 - 5000$A range. Here, however, the performance is much
   superior to our earlier estimates.

\section{CONCLUSIONS AND FUTURE WORK}

MSE is on-track to be an exceptionally sensitive facility that can
probe the spectra of the faintest objects in the sky. Even in the
areas in which it currently formally fails to meet requirements, its
estimated performance is more sensitive than any other existing
large aperture multi-object spectroscopic system. For the NIR, we are
currently investigating operating the H-band at lower temperatures,
and also revisiting our models of the NIR sky; the latter is critical
to understand correctly since sky photons are a major source of noise
that limits our performance. At high resolution, new versions of the
spectrograph design are being considered to increase the throughput of
this challenging subsystem, including adopting off-axis collimators
and reducing the fiber size, which then eases some of the stresses on
the grating.

At the end of the Conceptual Design Phase, we estimate that MSE is very close to
meeting the very demanding sensitivity requirements, and we are
currently investigating design optimisations to reconcile those areas
where the estimated performance does not meet the requirements. The
proposed changes to the baseline design of MSE outlined in this paper
are being discussed in the Project Office and we expect many of these
recommendations will be adopted following full trade studies considering
other aspects of system design not described here (cost, risk, etc).

\acknowledgments % equivalent to \section*{ACKNOWLEDGMENTS}       
 
% References

\end{document}